**EXTENDED ABSTRACT**

**EVALUATING THE EFFECT OF GEOMETRY AND CONTROL ON FREEWAY
MERGE BOTTLENECK CAPACITY**

**Mohamadamin Asgharzadeh, EIT**
Graduate Research Assistant
Department of Civil, Environmental and Architectural Engineering
2160A Learned Hall, 1530 W. 15th Street
University of Kansas, Lawrence, KS 66045
Tel: 785-727-6314; Email: asgharzadeh@ku.edu

**Alexandra Kondyli, PhD, Corresponding Author**
Assistant Professor
Department of Civil, Environmental and Architectural Engineering
2159A Learned Hall, 1530 W. 15th Street
University of Kansas, Lawrence, KS 66045
Tel: 785-864-6521; Email: akondyli@ku.edu

Word count: 1745 + 2 x Tables + 2 x Figures

Revised Submission Date: November 3, 2018

## ABSTRACT

This paper evaluates the impact of geometric and operational features on freeway capacity at merge bottleneck locations, by analyzing pre-breakdown flow rates and using parametric and non-parametric techniques. These features include number of lanes, free-flow speed or speed limit, length of acceleration lane, and presence or ramp metering. The analysis was carried out on seventeen freeway merge sites across the United States. The number of lanes showed negative relationship with the per-lane average pre-breakdown flow rate. Sites equipped with ramp meters also showed higher pre-breakdown flow rates than unmetered sites. In survival analysis, the number of lanes and presence of ramp meters were the only features found to have a statistically significant impact on the survival (or breakdown) probability. Capacities, defined at the flow rate that corresponds to 15 percent breakdown probability, were 2,048, 1,959, and 1,745 pc/hr/ln at three-lane, four-lane, and five-lane unmetered sites, respectively. Capacities were 2,248 pc/hr/ln at three-lane and 2,132 pc/hr/ln at four-lane metered sites. The effect of acceleration lane length and Free Flow Speed (FFS) on pre-breakdown flow rate and survival probability was not found to be significant.

## INTRODUCTION

The objective of this research is to investigate the impact of geometric and operational parameters on merge capacity, by analyzing data from various sites across the U.S. Flow characteristics such as heavy vehicles, FFS, presence of high-occupancy vehicle (HOV) or high-occupancy toll (HOT) lanes, ramp meters, and variable speed limit (VSL) are expected to affect freeway capacity. In addition, capacity varies from day to day even for the same site, therefore the stochastic nature of capacity should also be considered.

Using macroscopic data, pre-breakdown flow analysis was initially performed and statistical tests were implemented to compare sites that feature similar operational or geometric characteristics. More specifically, the number of mainline lanes, the acceleration lane length, the free flow speed, and the presence of ramp meters were considered as the basis of comparison. To account for the stochastic nature of capacity, lifetime survival analysis was also performed. The analysis included both non-parametric and parametric survival analysis. Survival functions were initially estimated exclusively for each site and then at aggregated level, based on common characteristics. Appropriate statistical tests were used to compare the survival functions. Recommendations regarding factors that affect merge capacities and their values are offered.

## METHODOLOGY

### Data Collection

Fourteen months of macroscopic data (from January 2017 to February 2018) including speeds, flow rates, and heavy vehicle percentage, were collected in 5-min intervals at seventeen merge bottlenecks across the U.S. California data were obtained from PeMS database (http://pems.dot.ca.gov), Kansas data were obtained from KC Scout (www.kcscout.com), and the Florida data were obtained from RITIS (https://ritis.org).

All sites were active merge bottlenecks, subject to recurring breakdown, with 12-ft lanes. Access to HOV/HOT lanes (if present) one mile upstream/downstream of the merge area was restricted (solid line or barrier separated). Also, sites had equal number of lanes downstream and upstream of the merge (i.e., no lane additions or drops).

Breakdowns were identified by monitoring the sensor located at the merge location. The following procedure were followed to identify true breakdowns and distinguish them from congestion due to downstream spillbacks:

1.   FFS (Table 1) was calculated as the volume-weighted average speed when flow rates were less than 1,000 pc/hr/ln and speeds were higher than 50 mph.

2.   A combined algorithm of speed drop (*1-3*) and breakdown speed threshold (*4-6*) was used to identify breakdowns. Sudden speed drop of 8% between two successive 5-min intervals at the vicinity of the breakdown speed threshold (75% of FFS) (*7*), was considered as a trigger to breakdown.

3.   During congestion, speeds remained below the breakdown speed threshold for at least 15 minutes. When speed exceeded the recovery speed (90% of FFS) (*7*), it was assumed that operations have recovered.

4.   Downstream spillback was addressed by monitoring sensors further downstream of the merge. In general, if a breakdown was observed at the downstream sensor at the same time or earlier than the breakdown recorded at the merge sensor, the breakdown at the merge sensor was assumed to be caused by the downstream queue propagation and the event was removed from further analysis (*8*).

5.   Observations with precipitation greater than 0.20 inches, identified from Weather Underground database (https://www.wunderground.com), were discarded.

6.   Incidents records downstream of the merge were retrieve from the respective databases, and discarded from the analysis dataset.

7.   Heavy traffic was converted into passenger cars (PC) using the passenger car equivalency equation provided in the HCM (*7*).

**Pre-Breakdown Flow and Survival Analysis**

Pre-breakdown flow analysis was initially performed to compare all sites. Pre-breakdown flow rate is the flow rate immediate prior to the breakdown event. Multiple factors were considered to compare sites at the aggregate level (based on common criteria), such as number of lanes (N), length of acceleration lane (ACC), ramp volume, and ramp metering operation.

The survival analysis requires both censored and uncensored data. Censored data include the pre-breakdown flow observations, while uncensored data represent all uncongested observations (*9*). All congested data points were removed from the analysis. Both censored and uncensored data were used to build the survival probability (and breakdown probability) function. The Kaplan-Meier (KME) survival and breakdown probability functions were calculated using Equations (1) and (2).

$$S_{KME}(q) = \prod_{i:q_i \leq q} \left(1 - \frac{d_i}{k_i}\right) \tag{1}$$

$$F_{KME}(q) = 1 - S_{KME}(q) \tag{2}$$

where,

$S_{KME}(q)$   = survival probability function,

$F_{KME}(q)$   = breakdown probability function,

$q$   = flow rate (pc/hr/ln),

$q_i$   = flow rate in interval $i$ (pc/hr/ln),

$k_i$    = number of intervals where $q_i \le q$, and
$d_i$    = number of breakdowns at volume $q_i$.

The survival functions were then compared using the Log-Rank non-parametric test (Mantel-Cox test) (*10*).

## Parametric Method

In parametric analysis, the Weibull distribution was fitted to the breakdown probability (*11*). The distribution function parameters were estimated by maximizing the Log-Likelihood Estimator (*2; 5; 12; 13*). The corresponding 15[th] and 20[th] percentile flow rates and the Weibull fitted parameters were chosen for comparison. Additionally, the optimum flow (*5; 12; 14*) was calculated using Equation (3).

$$q_{opt} = \beta \left(1/\alpha\right)^{1/\alpha} \tag{3}$$

where, $\beta$ and $\alpha$ are scale and shape of the Weibull distribution, respectively.

## FINDINGS

### Pre-Breakdown Flow Analysis

Each dataset was initially tested for Normality, using the Shapiro-Wilk test (*15*) and the assumption of homogeneity was assessed by Levene's Test (*16*). Since the Normality and Homogeneity assumptions were not met, Mann-Whitney (Wilcoxon signed rank test) (*17*) unpaired double-comparison test were utilized due to its robustness to these violations (Table 1). A 95% confidence level was chosen as the basis of comparison. For the non-parametric analysis, survival functions were constructed for unmetered and metered sites with different number of lanes. To assert whether the survival functions differ from each other, the Log-rank test was used (Table 1).

Based on Table 1, three-lane unmetered sites have higher pre-breakdown flow rate than four-lane and five-lane sites by 107.6 and 269.3 pc/hr/ln, respectively. At metered sites, the pre-breakdown flow rate at three-lane sites is higher by 246.1 compared to four-lane sites. While comparing the effect of the number of lanes, the pre-breakdown flow at three-lane and four-lane metered sites were higher by 188.1 and 246.0 pc/hr/ln than unmetered three-lane and four-lane sites, respectively. ACC and FFS were not found to contribute to statistically significant differences in the pre-breakdown flow rates between the sites. As for non-parametric survival analysis, the curves differ significantly by the number of lanes. Additionally, aggregate survival functions at metered and unmetered sites were statistically different. However, grouping the sites by FFS or ACC did not yield statistically different survival functions.

**TABLE 1 Pre-Breakdown Flow and Survival Analysis Results**

| Comparison | Common Characteristics | p-value | Diff * | 95% CI* | Inference* |
|---|---|---|---|---|---|
| **Pre-Breakdown Flow Analysis** | | | | | |
| 3-Lane and 4-Lane | Unmetered | <0.001 | 107.6 | [83.6  131.0] | Median at 4-lane sites is **lower by 107.7** than 3-lane sites. |
| 3-Lane and 5-Lane | | <0.001 | 269.3 | [246.4  291.6] | Median at 5-lane sites is **lower by 269.3** than 3-lane sites. |
| 4-Lane and 5-Lane | | <0.001 | 156.5 | [142.1  170.9] | Median at 5-lane sites is **lower by 156.5** than 4-lane sites. |
| ACC < 900 and ACC ≥ 900 ft | | 0.081 | NA | NA | Median difference is **not statistically significant**. |
| 3-Lane, 4-Lane | Metered | <0.001 | 246.1 | [226.0  266.1] | Median at 4-lane sites is **lower by 246.1** than three-lane sites. |
| Metered and Unmetered | 3-Lane | <0.001 | 188.1 | [160.0  216.9] | Median at three-lane metered sites is **greater by 188.1** than three-lane unmetered sites |
| Metered and Unmetered | 4-Lane | <0.001 | 246.0 | [223.2  267.3] | Median at four-lane metered sites is **greater by 246.0** than four-lane unmetered sites. |
| FFS<68 and FFS>68 mph | Unmetered 4-Lane | 0.670 | NA | NA | Median difference is **not statistically significant**. |
| **Survival Analysis Log-Rank Test** | | | | | |
| Number of lanes | Unmetered | < 0.001 | NA | NA | Survival functions **are significantly different** |
| Number of lanes | Metered | < 0.001 | NA | NA | Survival functions **are significantly different** |
| Number of lanes | Metered | < 0.001 | NA | NA | Survival functions **are significantly different** |
| Ramp meter | - | < 0.001 | NA | NA | Survival functions **are significantly different** |
| Free flow speed | Unmetered 4-lane | 0.20 | NA | NA | Survival functions **are not statistically different** |
| Length of acceleration lane | Unmetered four-lane | 0.10 | NA | NA | Survival functions **are not significantly different** |

*Units in pc/hr/ln

## Parametric Survival Analysis

Table 2 presents the Weibull distribution parameters (shape and scale), the resulting 15th and 20th percentile flow rates, and the optimum flow rate for each site. Linear trend lines were introduced for metered and unmetered sites with different number of lanes along with the respective R-square values. The number of lanes had negative association with the scale parameter at both metered and unmetered sites (Figure 1a). The 15th percentile flow rate also showed a decreasing trend as the number of lanes increased (Figure 1c). The Weibull shape parameter had a negative relationship with the number of lanes at metered sites while the trend was positive at unmetered sites (Figure 1b). Finally, the optimum flow showed a consistent negative association with the number of lanes at both metered and unmetered sites (Figure 1d).

All data from sites with common characteristics (number of lanes and ramp meter) were aggregated and breakdown probability functions were built for each group using the Product-Limit Method (PLM) (Figure 2).

Even at the aggregate level, metered sites showed higher 15th and 20th percentile flow rates than unmetered sites. As number of lanes dropped from five to three at unmetered sites, the optimum flow rate decreased from 1,931 pc/hr/ln to 1,649 pc/hr/ln. Similarly, the optimum flow rate at metered sites was 2,118 pc/hr/ln and 2,009 pc/hr/ln for three-lane and four-lane sites, respectively.

**TABLE 2 Parametric Weibull Distribution**

| ID | Weibull* (Shape) | Weibull* (Scale) | Lanes | Ramp Meter | 15th Percentile Flow Rate* | 20th Percentile Flow Rate* | Optimum Flow Rate* |
|---|---|---|---|---|---|---|---|
| **Individual Sites** | | | | | | | |
| 1 | 22.28 | 1,812 | 5 | N | 1,670 | 1,694 | 1,576 |
| 2 | 23.45 | 2,344 | 4 | Y | 2,169 | 2,199 | 2,049 |
| 3 | 18.30 | 2,121 | 4 | N | 1,920 | 1,955 | 1,809 |
| 4 | 23.95 | 1,949 | 4 | N | 1,806 | 1,831 | 1,707 |
| 5 | 20.81 | 1,984 | 4 | N | 1,818 | 1,846 | 1,715 |
| 6 | 20.77 | 2,191 | 4 | N | 2,008 | 2,038 | 1,893 |
| 7 | 15.29 | 2,285 | 4 | Y | 2,029 | 2,072 | 1,912 |
| 8 | 25.21 | 2,489 | 3 | Y | 2,316 | 2,340 | 2,190 |
| 9 | 22.09 | 1,957 | 5 | N | 1,802 | 1,829 | 1,701 |
| 10 | 16.28 | 2,259 | 4 | N | 2,020 | 2,060 | 1,903 |
| 11 | 16.16 | 2,265 | 3 | N | 2,024 | 2,064 | 1,907 |
| 12 | 28.21 | 1,863 | 5 | N | 1,747 | 1,766 | 1,655 |
| 13 | 23.63 | 2,215 | 3 | N | 2,051 | 2,079 | 1,938 |
| 14 | 22.61 | 2,241 | 3 | Y | 2,068 | 2,098 | 1,952 |
| 15 | 20.15 | 2,193 | 4 | N | 2,004 | 2,036 | 1,889 |
| 16 | 27.57 | 2,094 | 4 | N | 1,961 | 1,983 | 1,857 |
| **Aggregated** | | | | | | | |
| I | 2,252.4 | 19.16 | 3 | N | 2,048 | 2,083 | 1,931 |
| II | 2,162.6 | 18.37 | 4 | N | 1,959 | 1,993 | 1,846 |
| III | 1,888.8 | 23.06 | 5 | N | 1,745 | 1,770 | 1,649 |
| IV | 2,467.8 | 19.44 | 3 | Y | 2,248 | 2,284 | 2,118 |
| V | 2,357.6 | 18.05 | 4 | Y | 2,132 | 2,170 | 2,009 |

* Units in pc/hr/ln

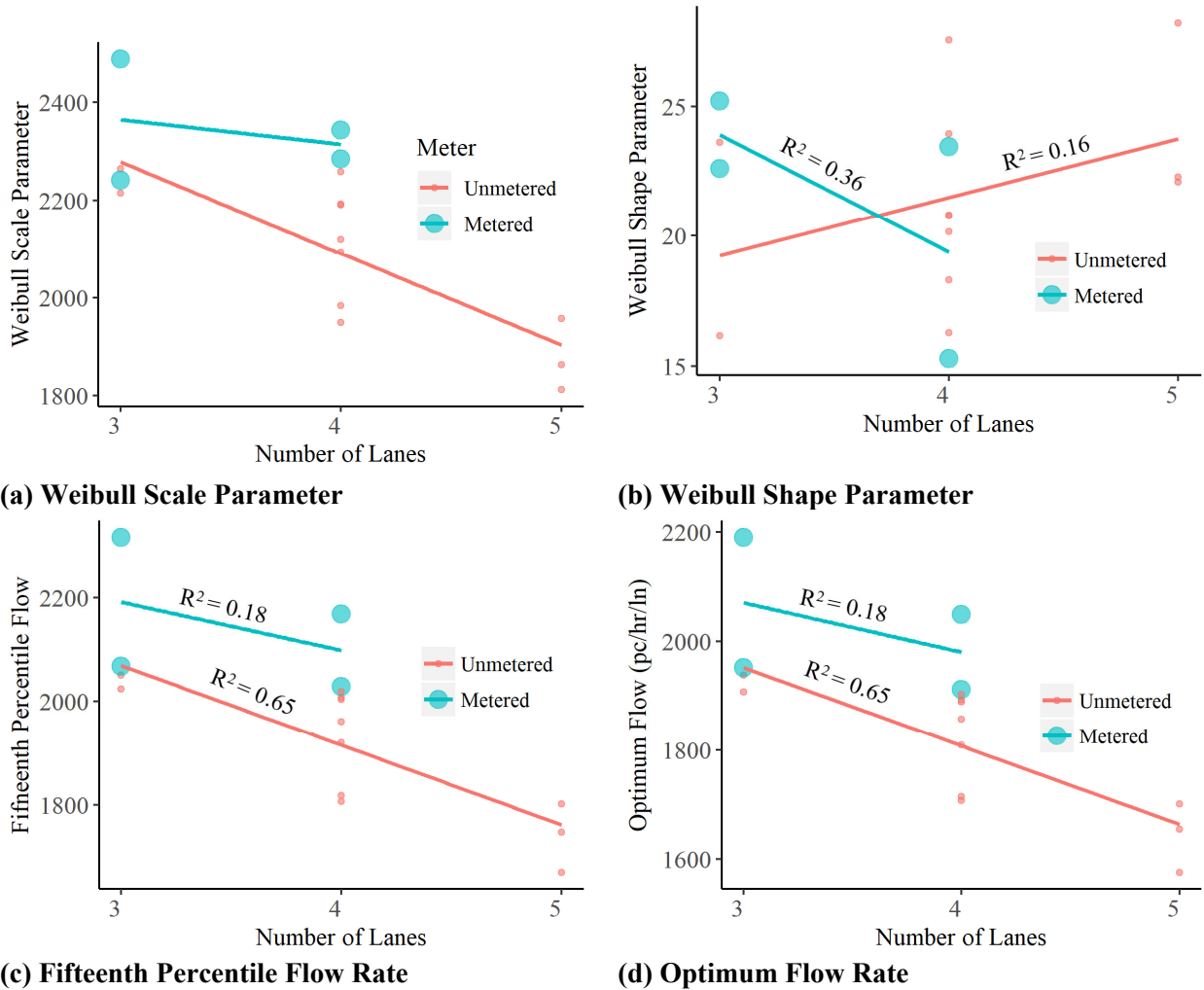

**(a) Weibull Scale Parameter**

**(b) Weibull Shape Parameter**

**(c) Fifteenth Percentile Flow Rate**

**(d) Optimum Flow Rate**

**FIGURE 1 Average Weibull Parameters (a) Scale, (b) Shape, (c) 15th Percentile Flow Rate, and (d) Optimum Flow Rate.**

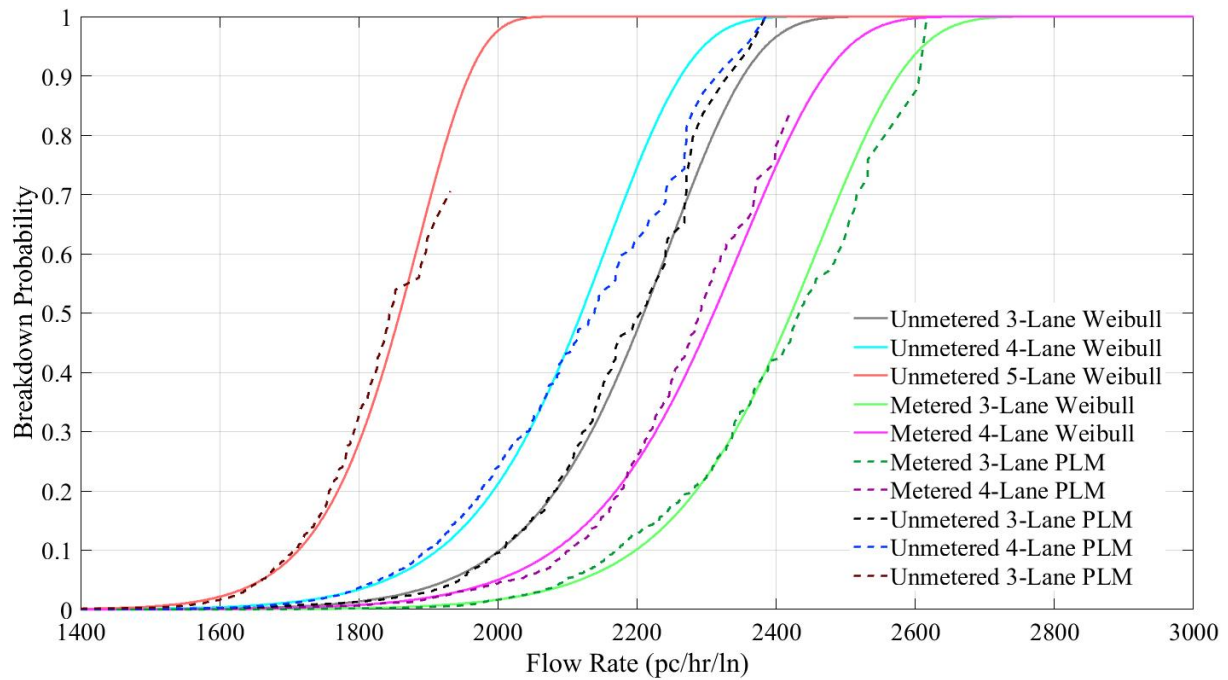

**FIGURE 2 PLM and Weibull Function.**

## CONCLUSIONS

This paper evaluated the effect of geometric and operational features on freeway capacity, using pre-breakdown flow rate analysis and non-parametric/parametric survival analysis. The results are summarized as below:

- **Pre-Breakdown Analysis**
1. The number of lanes showed negative relationship with the average pre-breakdown flow rate at all sites. At unmetered sites, the median three-lane pre-breakdown flow rate was greater by 107.7 pc/hr/ln and 269.3 pc/hr/ln than the four-lane and five-lane sites, respectively. At metered sites, the median three-lane pre-breakdown flow rate was greater than the four-lane sites by 246.1 pc/hr/ln.
2. The acceleration lane length and free flow speed did not have significant effect on the pre-breakdown flow rate.
3. Three-lane and four-lane sites with ramp meters had higher median pre-breakdown flow rate by 188.1 and 246.0 pc/hr/ln, compared to three-lane and four-lane sites without ramp metering, respectively.

- **Non-Parametric Survival Analysis**
1. Survival functions at sites with different number of lanes were statistically different from each other (at both metered and unmetered sites).
2. Aggregate survival functions were statistically different at metered and unmetered sites having the same number of lanes.
3. FFS and acceleration lane length did not have statistically significant impact on the survival function of four-lane unmetered sites.

- **Parametric Analysis**
1. Freeway capacity (15[th] percentile flow rate) had negative relationship with the number of lanes at both metered and unmetered sites. Capacity values at unmetered sites were 2,048, 1,959, and 1,745 pc/hr/ln at three-lane, four-lane, and five-lane sites, respectively. At metered sites, capacity values were 2,248 pc/hr/ln at three-lane sites and 2,132 pc/hr/ln at four-lane sites.
2. The optimum flow rate had a negative relationship with number of lanes.